\documentclass[acmsmall]{acmart}

\AtBeginDocument{%
  \providecommand\BibTeX{{%
    \normalfont B\kern-0.5em{\scshape i\kern-0.25em b}\kern-0.8em\TeX}}}

\usepackage{multirow} 
\usepackage{makecell} 

\setcopyright{acmcopyright}
\acmJournal{PACMHCI}
\acmYear{2020} \acmVolume{4} \acmNumber{CSCW1} \acmArticle{22} \acmMonth{5} \acmPrice{15.00}\acmDOI{10.1145/3392826}

\acmBooktitle{CSCW '20: ACM Conference on Computer Supported Cooperative Work and Social Computing}

\begin{document}

\title{How do Data Science Workers Collaborate? Roles, Workflows, and Tools}

\author{Amy X. Zhang}
\authornote{Both authors contributed equally to this research.}
\email{axz@cs.uw.edu}

\affiliation{%
  \institution{University of Washington \& }
  \institution{MIT}
  \country{USA}
}
\author{Michael Muller}
\authornotemark[1]
\email{michael_muller@us.ibm.com}
\affiliation{%
  \institution{IBM Research}
  \country{USA}
}

\author{Dakuo Wang}
\email{dakuo.wang@ibm.com}
\affiliation{%
  \institution{IBM Research \& }
  \institution{MIT-IBM Watson AI Lab}
  \country{USA}
  }








\begin{abstract}
Today, the prominence of data science within organizations has given rise to teams of data science workers collaborating on extracting insights from data, as opposed to individual data scientists working alone.
However, we still lack a deep understanding of how data science workers collaborate in practice. In this work, we conducted an online survey with 183 participants who work in various aspects of data science. 
We focused on their reported interactions with each other (e.g., managers with engineers) and with different tools (e.g., Jupyter Notebook).
We found that data science teams are extremely collaborative and work with a variety of stakeholders and tools during the six common steps of a data science workflow (e.g., clean data and train model).
We also found that the collaborative practices workers employ, such as documentation, vary according to the kinds of tools they use. 
Based on these findings, we discuss design implications for supporting data science team collaborations and future research directions.
\end{abstract}



\maketitle

\section{Introduction}
Data science often refers to the process of leveraging modern machine learning techniques to identify insights from data~\cite{kim2016emerging,muller2019datascience,kaggle2018survey}. In recent years, with more organizations adopting a ``data-centered'' approach to decision-making~\cite{ufford2018beyond,dobrin_ibm_analytics_2017}, more and more \textit{teams} of data science workers have formed to work \textit{collaboratively} on larger data sets, more structured code pipelines, and more consequential decisions and products. Meanwhile, research around data science topics has also increased rapidly within the HCI and CSCW community in the past several years~\cite{guo2011proactive,kim2016emerging,muller2019datascience,kross2019practitioners,rule2018exploration,kery2018story,wang2019humanai,wang2019data,wang2020callisto}.

From existing literature, we have learned that the data science workflow often consists of multiple phases~\cite{muller2019datascience,wang2019humanai,kross2019practitioners}. For example, Wang et al.~describes the data science workflow as containing 3 major phases---Preparation, Modeling, and Deployment---and 10 more fine-grained steps~\cite{wang2019humanai}. 
Various tools have also been built for supporting data science work, including programming languages such as Python or R, statistical analysis tools such as SAS~\cite{r39} and SPSS~\cite{r43}, integrated development environments (IDEs) such as Jupyter Notebook~\cite{kluyver2016jupyter,granger2017jupyterlab}, and automated model building systems such as AutoML~\cite{web:googleautoml,liu2019admm} and AutoAI~\cite{wang2019humanai}. 
And from empirical studies, we know how individual data scientists are using these tools~\cite{kery2018story,kery2019towards,rule2018exploration}, and what features could be added to improve the tools for users working alone~\cite{wang2019data}. 

However, a growing body of recent literature has hinted that data science projects consist of complex tasks that require \textit{multiple} skills~\cite{kim2016emerging, r42}. 
These requirements often lead participants to juggle multiple roles in a project, or to work in teams with others who have distinct skills.
For instance, in addition to the well-studied role of data scientist~\cite{r37, r44}, who engages in technical activities such as
cleaning data, extracting or designing features, analyzing/modeling data, and evaluating results, there is also the role of project manager, who engages in less technical activities such as reporting results~\cite{hou2017hacking, r36, r51, r52}. The 2017 Kaggle survey reported additional roles involved in data science \cite{hayes2018}, but without addressing topics of collaboration. In this work, we limit the roles in our survey to activities and relationships that were mention in interviews in Muller et al.~\cite{muller2019datascience}

Unfortunately, most of today's understanding of data science collaboration only focuses on the perspective of the data scientist, and how to build tools to support distant and asynchronous collaborations among data scientists, such as version control of code.
The technical collaborations afforded by such tools~\cite{r70} only scratch the surface of the many ways that collaborations may happen within a data science team, such as when stakeholders discuss the framing of an initial problem before any code is written or data collected~\cite{pine2015politics}. 
However, we have little empirical data to characterize the many potential forms of data science collaboration.

Indeed, we should not assume data science team collaboration is the same as the activities from a conventional software
development team, as argued by various previous literature~\cite{guo2011proactive,kross2019practitioners}. 
Data science is engaged as an ``exploration'' process more than an ``engineering''
process~\cite{r25, kery2018story, muller2019datascience}. 
``Engineering'' work is oftentimes assumed to involve extended blocks of solitude, without the benefit of colleagues' expertise while engaging with data and code~\cite{parnin2013programmer}. 
While this perspective on engineering is still evolving \cite{storey2006shared}, there is no doubt that ``exploration'' work requires deep domain knowledge that oftentimes only resides in domain experts' minds~\cite{kery2018story,muller2019datascience}. 
And due to the distinct skills and knowledge residing within different roles in a data science team, more challenges with collaboration can arise~\cite{mao2019,hou2017hacking}.

In this paper, we aim to deepen our current understanding of the collaborative practices of data science teams from not only the perspective of technical team members (e.g., data scientists and engineers), but also the understudied perspective of non-technical team members (e.g., team managers and domain experts). Our study covers both \textbf{a large scale of users}---we designed an online survey and recruited 183 participants with experience working in data science teams---and \textbf{an in-depth investigation}---our survey questions dive into 5 major roles in a data science team (engineer/analyst/programmer, researcher/scientist, domain expert, manager/executive, and communicator), and 6  stages (understand problem and create plan, access and clean data, select and engineer features, train and apply models, evaluate model outcomes, and communicate with clients or stakeholders) in a data science workflow. In particular, we report what other roles a team member works with, in which step(s) of the workflow, and using what tools.

In what follows, we first review literature on the topic of data science work practices and tooling; then we present the research method and the design of the survey; we report survey results following the order of \emph{Overview of Collaboration, Collaboration Roles, Collaborative Tools, and Collaborative Practices}. Based on these findings, we discuss implications and suggest designs of future collaborative tools for data science. 

\section{Related Work}
Our research contributes to the existing literature on how data science teams work.
We start this section by reviewing recent HCI and CSCW research on data science work practices; then we take an overview of the systems and features designed to support data science work practices. Finally, we highlight specific literature that aims to understand and support particularly the collaborative aspect of data science teamwork. 

\subsection{Data Science Work Practices}
Jonathan Grudin describes the current cycle of the popularity of AI-related topics in industry and in academia as an ``AI Summer''~\cite{grudin2009ai}.
In the hype surrounding AI, many fancy technology demos mark key milestones, such as IBM DeepBlue~\cite{campbell2002deep}, which defeated a human chess champion for the first time, and Google's AlphaGo demo, which defeated the world champion in Go~\cite{wang2016does}. With these advances in AI and machine learning technologies, more and more organizations are trying to apply machine learning models to business decision-making processes. People refer to this collection of work as ``data science''~\cite{guo2011proactive,kross2019practitioners,kaggle2018survey,muller2019datascience} and the various workers who participate in this process as ``data scientists'' or ``data engineers''. 

\begin{figure}[ht]
  \begin{center}
    \includegraphics[width=\columnwidth]{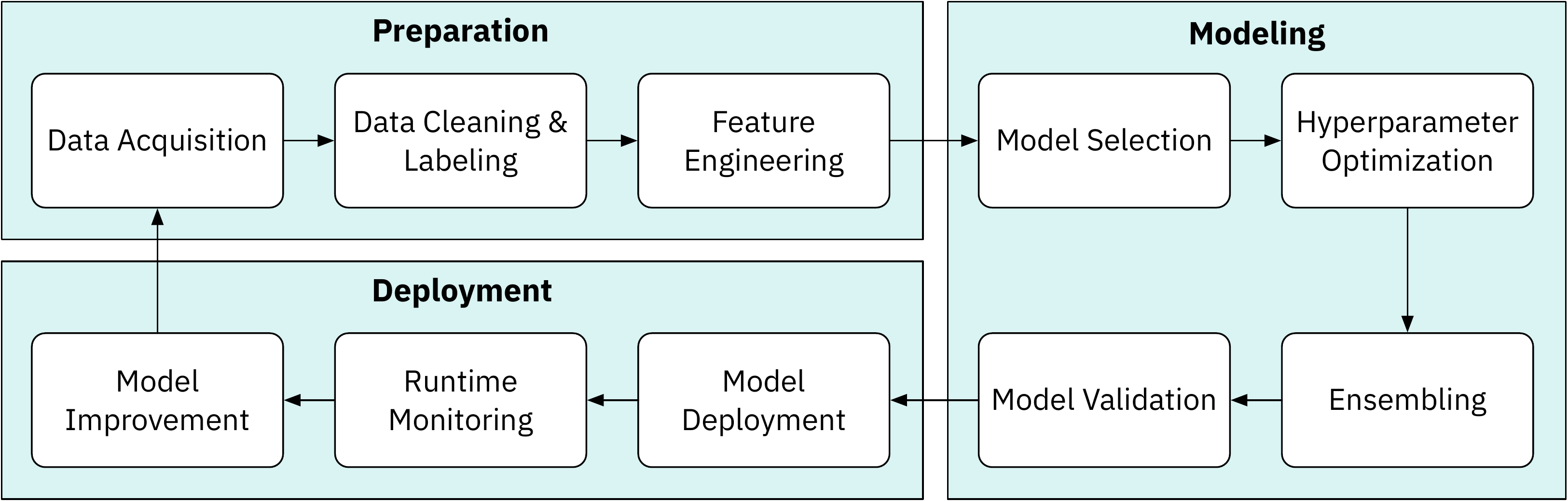}
  \end{center}
  \caption{A data science workflow, consisting of three high-level phases: data preparation, model building, and model deployment ~\cite{wang2019humanai} }
  \label{fig:DS-steps}
\end{figure}

HCI researchers are interested in data science practices. Studies have been conducted to understand data science work practices~\cite{passi2017data,passi2018trust,guo2011proactive,kross2019practitioners,kery2018story,rule2018exploration,hou2017hacking,mao2019,muller2019datascience}, sometimes using the label of Human Centered Data Science \cite{aragon2016developing, muller2019human}. For example, Wang et al.~proposed a framework of 3 stages and 10 steps to characterize the data science workflow by synthesizing existing literature (Figure~\ref{fig:DS-steps})~\cite{wang2019humanai}. The stages consist of Preparation, Modeling, and Deployment, and at a finer-grained level, the framework has 10 steps from Data Acquisition to Model Runtime Monitoring and Improvement. This workflow framework is built on top of Muller et al.'s work~\cite{muller2019datascience}, which mostly focused on the Preparation steps, and decomposed the data science workflow into 4 stages, based on interviews with professional data scientists: Data Acquisition, Data Cleaning, Feature Engineering, and Model Building and Selection.

Given these workflow frameworks and terminology~\cite{muller2019datascience,wang2019humanai}, we can position existing empirical work within a data science workflow. For example, researchers have suggested that 80\% of a data science project is spent in the Preparation stage ~\cite{zoller2019survey,muller2019datascience,guo2011proactive,kandel2011wrangler, rattenbury2017principles}. 
As a result, data scientists often do not have enough time to complete a comprehensive data analysis in the Modeling stage~\cite{sutton2018data}.
Passi and Jackson dug deeper into the Preparation stage, showing that when data scientists pre-process their data, it is often a rule-based, not rule-bound, process~\cite{passi2017data}. 
Pine and Liboiron further investigated how data scientists made those pre-processing rules~\cite{pine2015politics}.

However, most of the literature focuses only on a single data scientist's perspective, despite many interviewees reporting that ``data science is a team sport''~\cite{wang2019humanai}. Even in the Muller et al.~\cite{muller2019datascience} and Wang et al.~\cite{wang2019humanai} workflows, they focus only on the activities that involve data and code, which were most likely performed by the technical roles in the data science team. 
The voices of the non-technical collaborators within a data science team are missing, including an understanding of who they worked with when, and what tools they used.

In contrast to the current literature in data science, software engineering has built a strong literature on collaborative practices in software development~\cite{treude2009empirical,kraut1995coordination,herbsleb2001empirical}, including in both open source communities~\cite{bird2008latent,dabbish2012social} and industry teams~\cite{begel2008effecting}. As teams working on a single code base can often be large, cross-site, and globally dispersed, much research has focused on the challenges and potential solutions for communication and coordination~\cite{herbsleb2001empirical}. These challenges can be exacerbated by cultural differences between team members of different backgrounds~\cite{halverson2006designing,huang2007cultural} and by the different roles of team members such as project manager~\cite{zhang2007managing} or operator~\cite{tessem2008cooperation}.
Many of the tools used by software engineering teams are also used by data science teams (i.e., GitHub~\cite{dabbish2012social}, Slack~\cite{Coding_team:Park:2018:PPL:3266037.3266098}), and the lessons learned from this work can inform the design of collaborative tools for data science. However, there are also important differences when it comes to data science in particular, such as the types of roles and technical expertise of data science collaborators as well as a greater emphasis on exploration, data management, and communicating insights in data science projects.

Many of the papers about solitary data science work practices adopted the interview research method~\cite{wang2019data,guo2011proactive,kross2019practitioners,kery2018story,wang2019humanai,muller2019datascience}. An interview research method is well-suited for the exploratory nature of these empirical works in understanding a new practice, but it also falls short in generating a representative and generalizable understanding from a larger user population. Thus, we decided to leverage a large-scale online survey to complement the existing qualitative narratives.

\subsection{Collaboration in Data Science}
Only recently have some CSCW researchers began to investigate the collaborative aspect of data science work~\cite{grappiolo2019semantic,passi2017data,stein2017make,borgman2012s,viaene2013data}. 
For example, Hou and Wang~\cite{hou2017hacking} conducted an ethnography study to explore collaboration in a civic data hackathon event where data science workers help non-profit organizations develop insights from their data. 
Mao et al.~\cite{mao2019} interviewed biomedical domain experts and data scientists who worked on the same data science projects. 
Their findings partially echo previous results ~\cite{muller2019datascience,guo2011proactive,kross2019practitioners} that suggest data science workflows have many steps. More importantly, their findings are similar to the software engineering work cited above~\cite{treude2009empirical,kraut1995coordination,herbsleb2001empirical}
opening the possibility that data science may also be a highly \textit{collaborative} effort where domain experts and data scientists need to work closely together to advance along the workflow. 

Researchers also observed challenges in collaborations within data science teams that were not as common in conventional software engineering teams. Bopp et al.~showed that ``big data'' could become a burden to non-profit organizations who lack staff to make use of those data resources \cite{bopp2017disempowered}.
Hou et al.~provided a possible reason---i.e., that the technical data workers ``speak a different language'' than the non-technical problem owners, such as a non-profit organization (NPO) client, in the civic data hackathon that they studied~\cite{hou2017hacking}. The non-technical NPO clients could only describe their business questions in natural language, e.g., ``why is this phenomenon happening?'' But data science workers do not necessarily know how to translate this business question into a data science question. A good practice researchers observed in this context was ``brokering activity'', where a special group of organizers who understand both data science and the context serve as translators to turn business questions into data science questions (see Williams' earlier HCI work on the importance of ``translators'' who understand and mediate between multiple domains~\cite{williams1993translation}). Also, once the data workers generated the results, the ``brokers'' helped to interpret the data science findings into business insights.

The aforementioned collaboration challenges~\cite{hou2017hacking} are not unique to the civic data hackathon context. Mao et al.~\cite{mao2019} interviewed both data scientists and bio-medical scientists who worked together in research projects. They found that these two different roles often do not have common ground about the project's progress. For example, the goal of bio-medical scientists is to discover new knowledge; thus, when they ask a research question, that question is often tentative. 
Once there is an intermediate result, bio-medical scientists often need to \emph{revise their research question} or \emph{ask a new question}, because their scientific journey is to ``ask the right question''.
However, the data scientists were focused on transferring a research question into a well-defined data science question so they could optimize machine learning models and increase performance. 
 The behavior of the bio-medical scientists was perceived by the data scientists as ``wasting our time'', as they had worked hard to ``find the answer to the question'' that later was discarded. 
 Mao et al.~argued that the constant re-calibration of common ground might help to ease tensions and support cross-discipline data science work.

These related projects focused only on a civic data hackathon~\cite{hou2017hacking} and on the collaborative projects between data scientists and bio-medical scientists in scientific discovery projects~\cite{mao2019}. Also, both of them used ethnographic research methods aiming for in-depth understanding of the context. 
In this work, we wanted to target a more commonly available scenario---data science teams' work practices in corporations---as this scenario is where most data science professionals work. We also want to gather a broader user perspective through the deployment of an online survey.

\subsection{Data Science Tools}
Based on the empirical findings and design suggestions from previous literature~\cite{hou2017hacking,mao2019,muller2019datascience,grappiolo2019semantic,passi2017data,stein2017make,borgman2012s,viaene2013data}, some designers and system builders have proposed human-in-the-loop design principles for science tools~\cite{amershi2019guidelines,amershi2011human,wang2019data,kery2018story,kery2019towards,gil2019towards}. For example, Gil et al.~surveyed papers about building machine learning systems and developed a set of design guidelines for building human-centered machine learning systems~\cite{gil2019towards}. Amershi et al.~in parallel reviewed a broader spectrum of AI applications and proposed a set of design suggestions for AI systems in general~\cite{amershi2019guidelines}. 

With these design principles and guidelines in mind~\cite{gil2019towards,amershi2019guidelines}, many systems and features have been proposed to support aspects of data science work practices. 
One notable system is Jupyter Notebook~\cite{web:jupyter} and its variations such as Google Colab~\cite{web:colab} and Jupyter-Lab~\cite{web:jupyterlab}. Jupyter Notebook is an integrated code development environment tailored for data science work practices. 
It has a graphical user interface that supports three key functionalities---coding, documenting a narrative, and observing execution results~\cite{kross2019practitioners}---that are central to data science work~\cite{kery2018story}. Moreover, the ability to easily switch between code and output cells allows data scientists to quickly iterate on their model-crafting and testing steps~\cite{kery2018story,muller2019datascience}.

However, only a few recent works have started to look at designing specific collaborative features to support data science teams beyond the individual data scientist's perspective~\cite{wang2019data,wang2020callisto,rule2018exploration,chang18,muller2018jupytercon}. For example, Jupyter Notebook's narrative cell feature is designed to allow data scientists to leave human-readable annotations so that when another data scientist re-uses the code, they can better understand it. However, Rule et al.~found a very low usage of these narrative cells (markdown cells) among a million Jupyter notebooks that they sampled from GitHub~\cite{rule2018exploration}. Data scientists were not writing their notebooks with a future collaborator or re-user in mind. 

More recently, Wang and colleagues at University of Michigan have examined how data science tools can better support collaboration. Their 2019 study~\cite{wang2019data} aimed to understand if the Jupyter Notebook had a new feature that allows multiple data scientists to synchronously write code (as many people do in Google Docs today~\cite{dakuo}), whether and how data scientists would use it for their collaboration. 
They found the proposed feature can encourage more exploration and reduce communication costs, while also promoting unbalanced participation and slacker behaviors. 
In their 2020 paper~\cite{wang2020callisto}, Wang et al. took up a related challenge, namely the documentation of informal conversations and decisions that take place during data science projects. Building on prior work~\cite{chang18,muller2018jupytercon,Coding_team:Park:2018:PPL:3266037.3266098}, they built Callisto, an integration of synchronous chat with a Jupyter notebook.
In tests with 33 data science practitioners, Wang et al.~showed the importance of automatic computation of the reference point in order to anchor chat discussion in the code. 

In sum, almost all of the proposed tools and features in data science focus only on the technical users' scenarios (e.g., data scientists and data engineers), such as how to better understand and wrangle data~\cite{heer2007voyagers,dang2018predict}, or how to better write and share code~\cite{wang2019data, wang2020callisto,rule2018exploration,kery2019towards}. 
In this work, we want to present an account that covers both the technical roles and the non-technical roles of a professional data science team in corporations, so that we can better propose design suggestions from a multi-disciplinary perspective.

\section{Method}

\subsection{Participants}
Participants were a self-selected convenience sample of employees in IBM who read or contributed to Slack channels about data science (e.g., channel-names such as ``deeplearning'', ``data-science-at-ibm'', ``ibm-nlp'', and similar). Participants worked in diverse roles in research, engineering, health sciences, management, and related line-of-business organizations. 

We estimate that the Slack channels were read by approximately 1000 employees. Thus, the 183 people who provided data constituted a 20\% percent participation rate.

Participants had the option to complete the survey anonymously. Therefore, our knowledge of the participants is derived from their responses to survey items about their roles on data science teams (Figure \ref{fig:experience}).

\subsection{Survey Questions}
The survey asked participants to describe a recent data science project, focusing on collaborations (if any), the roles and skills among the data science team (if appropriate), and the role of collaborators at different stages of the data science workflow.
Next, we asked open-ended questions about the tools participants used to collaborate, including at different workflow stages.\footnote{Open-text responses were coded by two of the authors. We agreed on a set of coding guidelines in advance, and we resolved any disagreements through discussion.}
Finally, we asked participants to describe their collaborative practices around sharing and re-using code and data, including their expectations around their own work and their documentation practices.
To encourage more people to contribute, we made all questions optional.

\subsection{Survey Distribution}
We posted requests to participate in relevant IBM internal Slack channels during January 2019.
Responses began to arrive in January. We wrote 2--4 reminder posts, depending on the size and activity of each Slack channel. We collected the last response on 3 April 2019.

\section{Results}

\begin{figure}
\includegraphics[width=1.0\columnwidth]{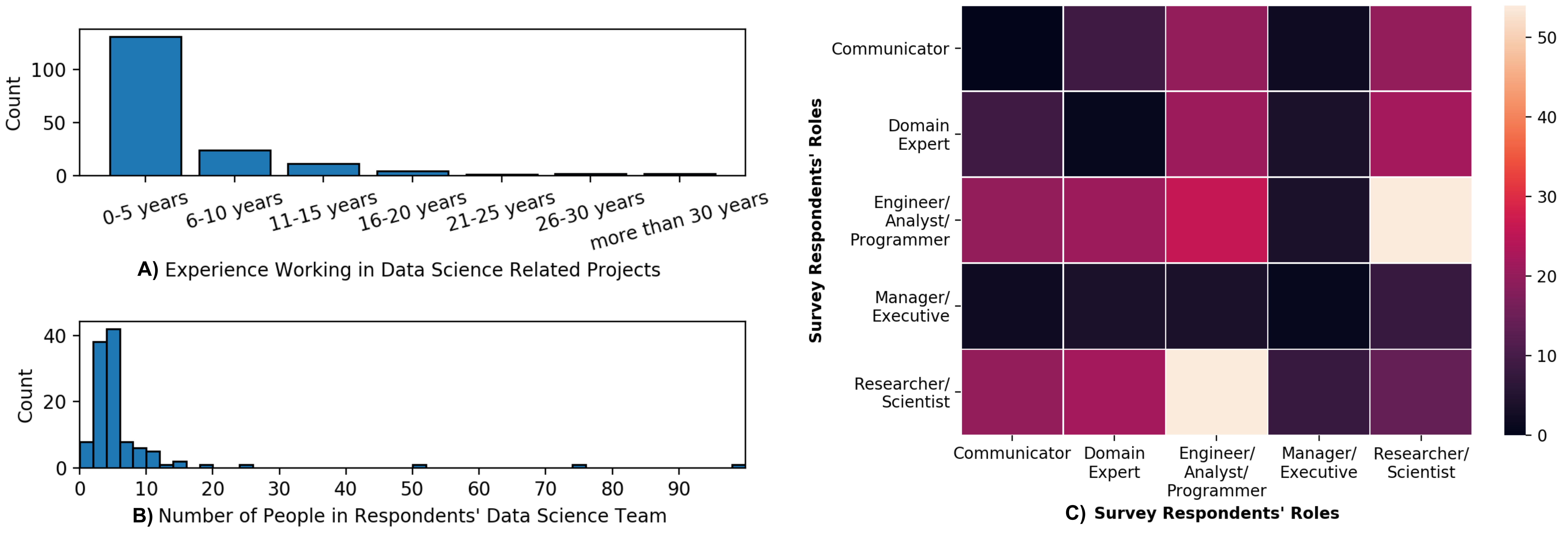}
\caption{ Self-reported information about survey respondents: A) Work experience in data science and machine learning-related projects. B) Histogram of data science team sizes. C) Heatmap of the prevalence of a pair of roles taken on by one respondent within in a data science project, with the diagonal showing the respondents who only self-identified as one role.}
\label{fig:experience}
\end{figure}



The 183 people who responded to the anonymous survey described themselves as being of varied experience in data science but primarily 0--5 years (Figure \ref{fig:experience}A). Clearly, some saw connections between contemporary data science and earlier projects involving statistical modeling, and that is why we see some long years of experience. Respondents worked primarily in smaller teams of six people or fewer (Figure \ref{fig:experience}B). A few appeared to have solo data science practices.

Respondents reported that they often acted in multiple roles in their teams, and this may be due to the fact that most of them have a relatively small team. Figure \ref{fig:experience}C is a heatmap showing the number of times in our survey a respondent stated they acted in both roles out of a possible pair (with the two roles defined by a cell's position along the x-axis and y-axis). For cells along the diagonal, we report the number of respondents who stated they only performed that one role and no other. As can be seen, this was relatively rare, except in the case of the Engineer/Analyst/Programmer role.

Unsurprisingly, there was considerable role-overlap among Engineers/Analysts/Programmers and Researchers/Scientists (i.e., the technical roles). These two roles also served---to a lesser extent---in the roles of Communicators and Domain Experts.

By contrast, people in the role of Manager/Executive reported little overlap with other roles. From the roles-overlap heatmap of Figure \ref{fig:experience}C, it appears that functional leadership---i.e., working in multiple roles---occurred primarily in technical roles (Engineer/Analyst/Programmer and Researcher/Scientist). These patterns may reflect IBM's culture to define managers as people-managers, rather than as technical team leaders.


\subsection{Do Data Science Workers Collaborate?} \label{do_collaborate}

Figure \ref{fig:roles} shows patterns of self-reported collaborations across different roles in data science projects. 
First, we begin answering one of the overall research questions: \textbf{What is the extent of collaboration on data science teams?}

\begin{table}[b]
  \caption{Percentages of collaborations reported by each role.}
  \small
    \centering
    \begin{tabular}{l c}
    \toprule
    Role & Percent Reporting Collaboration\\
    \midrule
    Engineer/Analyst/Programmer & 99\%\\
    Communicator & 96\%\\
    Researcher/Scientist & 95\%\\
    Manager/Executive & 89\%\\
    Domain Expert & 87\%\\ 
    \bottomrule
    \end{tabular}
    \label{tab:collab_percent}
\end{table}

\subsubsection{Rates of Collaboration}
\begin{figure*}[]
\includegraphics[width=1\linewidth]{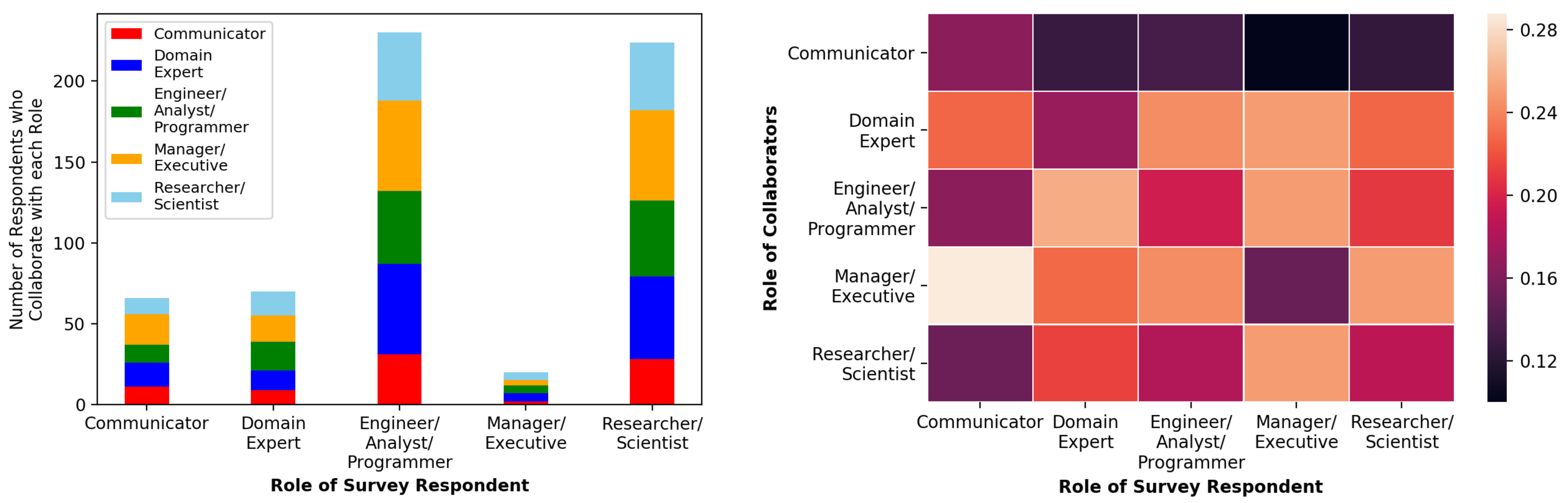}
\caption{Who collaborates with whom?  Note: raw counts are reported in the stacked bar chart and normalized percentages (along the columns) are reported in the heatmap.}
\label{fig:roles}
\end{figure*}

The data behind Figure \ref{fig:roles} allow us to see the extent of collaboration for each self-reported role among the data science workers (Table \ref{tab:collab_percent}). Among the five data science roles of Figure \ref{fig:roles}, three roles reported collaboration at rates of 95\% or higher. The lowest collaboration-rate was among Domain Experts, who collectively reported a collaboration percentage of 87\%. 
In the following subsections, we explore the patterns and supports for these collaborations.

\subsubsection{Who Collaborates with Whom?}
\label{sec:collaboration}
The stacked bar chart to the left in Figure \ref{fig:roles} reflects the raw numbers of people in each role who responded to our survey and stated that they collaborated with another role. 
The heatmap to the right of Figure \ref{fig:roles} shows a similar view of the collaboration relationship---with whom they believe they collaborate---as the chart on the left, except that the cells are now normalized by the total volume in each column.
The columns (and the horizontal axis) represent the \textit{reporter} of a collaborative relationship. The rows (and the vertical axis) represent the \textit{collaboration partner} who is mentioned by the reporter at the base of each column.
Lighter colors in the heatmap indicate more frequently-reported collaboration partnerships. 

When we examine a square heatmap with reciprocal rows and columns, we may look for asymmetries around the major diagonal. For each pair of roles (A and B), do the informants report a similar proportion of collaboration in each direction---i.e., does A report about the same level of collaboration with B, as B reports about A?

Surprisingly, we see a disagreement about collaborations in relation to the role of Communicator. Communicators report strong collaborations with Managers and with Domain Experts, as shown in the Communicator column of Figure \ref{fig:roles}. However, these reports are not fully reciprocated by those collaboration partners. As shown in the row for Communicators, most roles reported little collaboration with Communicators relative to the other roles. A particularly striking difference is that Communicators report (in their column) relatively strong collaboration with Managers/Executives, but the Managers/Executives (in their own column) report the least collaboration with Communicators. There is a similar, less severe, asymmetry between Communicators and Domain Experts. We will later interpret these findings in the Discussion in Section~\ref{sec:discussion_collaboration}.



\subsubsection{Are there ``Hub'' Collaborator Roles?}
\label{sec:intra_collaboration}

\begin{figure*}[t]
\includegraphics[width=0.5\linewidth]{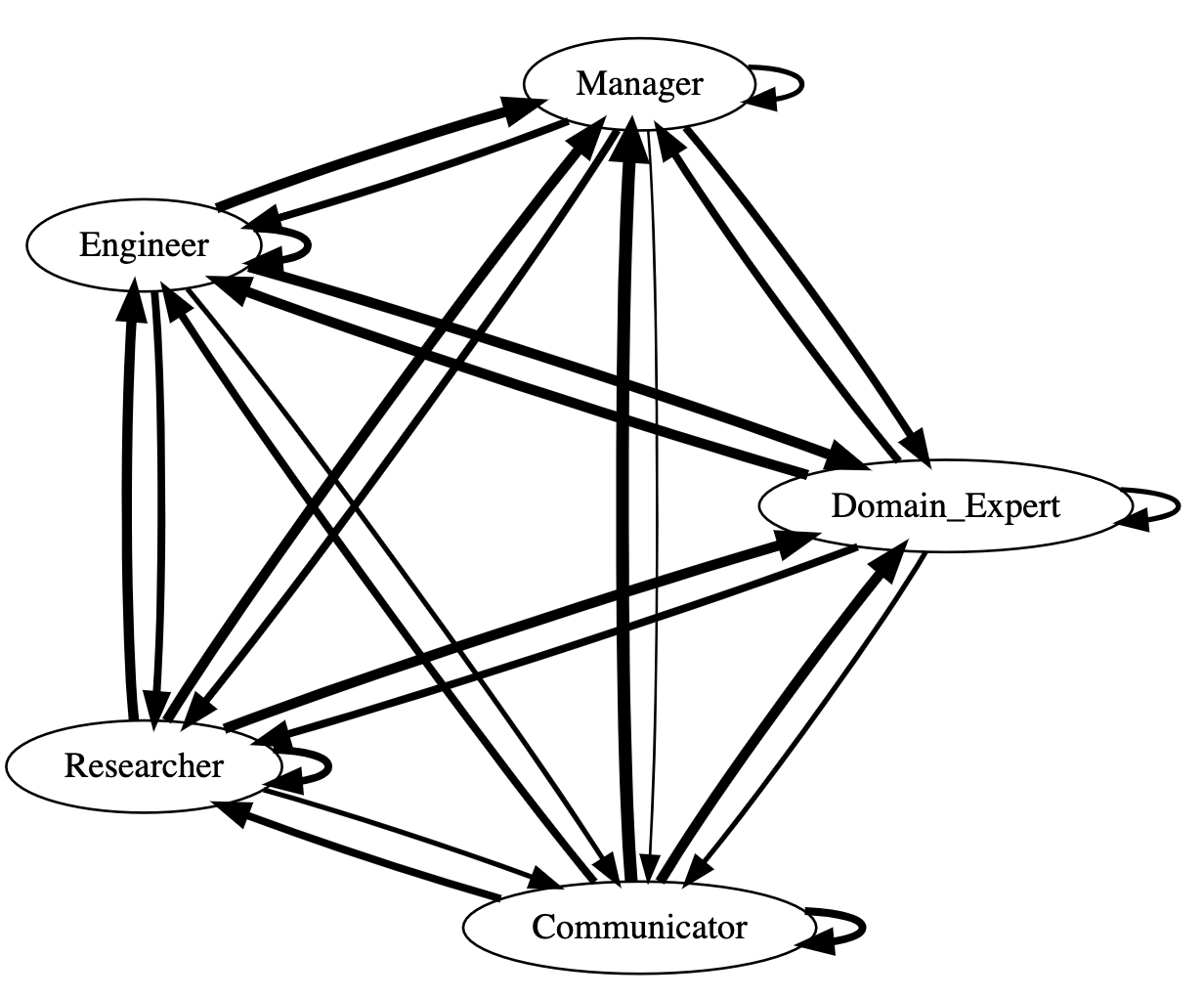}
\caption{Network graph of reported collaborative relationships. The arrow \textit{from} Researcher \textit{to} Communicator may be interpreted as aggregated reports \textit{by} Researchers about their collaboration \textit{with} Communicators. Note that some pairwise relationships do not have equal bidirectional symmetry. The thickness of each arc represents the proportion of people who reported each directed-type of collaboration, normalized by number of people in each role.}
\label{fig:collaborator_network}
\end{figure*}

Are certain roles dominant in the collaboration network of Figure \ref{fig:roles}? 
Figure \ref{fig:collaborator_network} shows the reports of collaboration from Figure \ref{fig:roles} as a network graph.
Each report of collaboration takes the form of a directed arc from one role to another. The direction of the arc between e.g. (A\-->B) can be interpreted as ``A reports collaboration with B.'' The thickness of each arc represents the proportion of people who report each directed-type of collaboration. To avoid distortions due to different numbers of people reporting in each role, we normalized the width of each arc as the number of reported collaborations divided by the number of people reporting from that role. Self-arcs represent cases in which the two collaborators were in the same role---e.g., an engineer who reports collaborating with another engineer.

With one exception, this view shows relatively egalitarian strengths of role-to-role collaboration. While we might expect to find Managers/Executives as the dominant or ``hub'' role, their collaborative relations are generally similar to those of Engineers and Researchers. Domain Experts are only slightly less engaged in collaborations.

The exception occurs, as noted above, in relation to Communicators. Communicators in this graph clearly believe that they are collaborating strongly with other roles (thick arrows), but the other roles report less collaboration toward Communicators (thin arrows).

The self-loop arrows are also suggestive. These arrows appear to show strong \emph{intra-role} collaborations among Engineers, Researchers, and Communicators. By contrast, Managers/Executives and Domain Experts appear to collaborate less with other members of their own roles.


\subsection{Collaborator Roles in Different Stages of the Data Science Workflow} \label{stages_roles}

As reviewed above in relation to Figure \ref{fig:DS-steps}, data science projects are often thought to follow a series of steps or stages---even if these sequences serve more as mental models than as guides to daily practice~\cite{muller2019datascience, passi2017data}. We now consider how the roles of data science workers interact with those stages.

\begin{figure*}[]
    \includegraphics[width=0.7\linewidth]{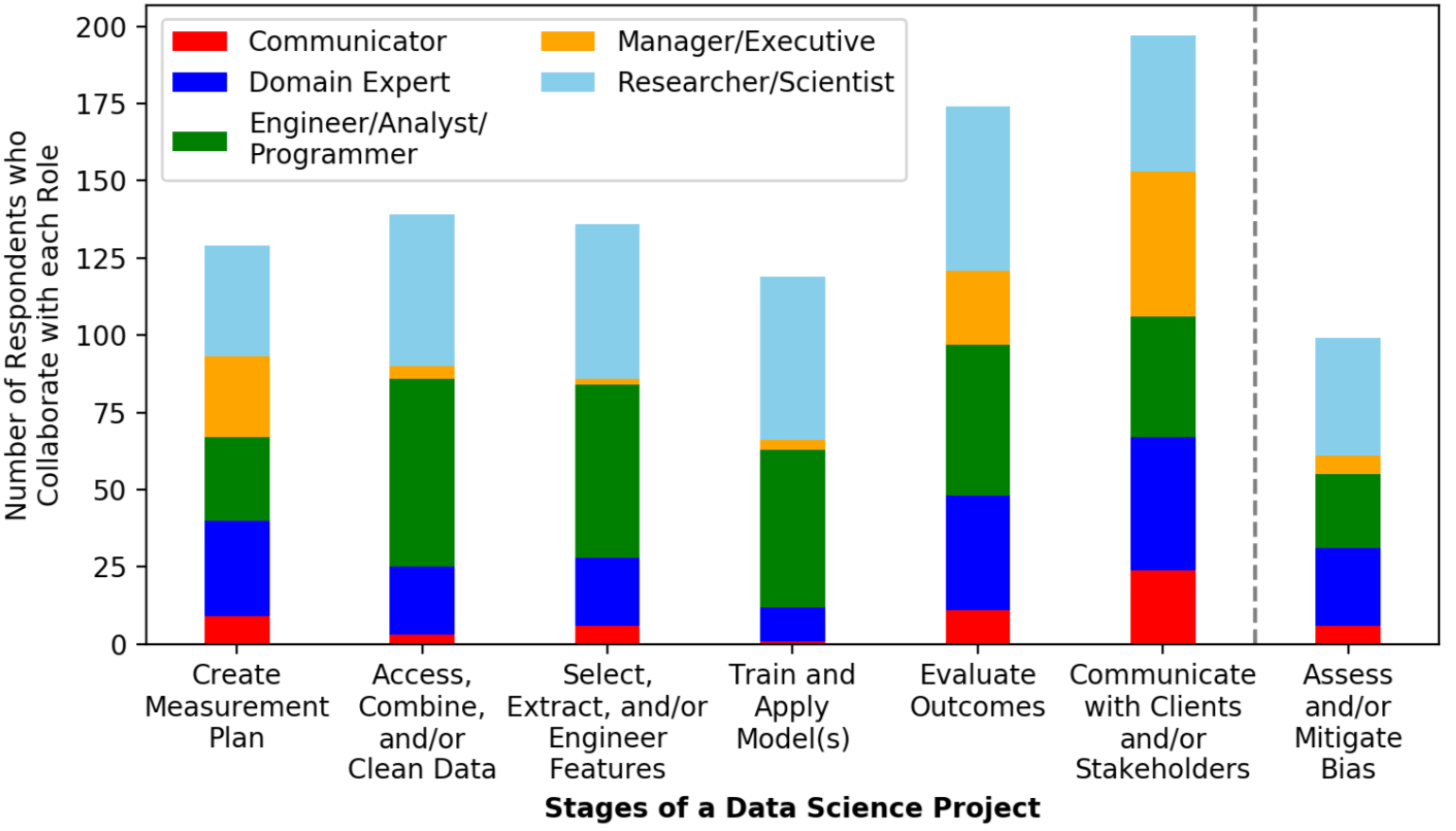}
\caption{The roles of collaborators during the stages of a data science project.}
\label{fig:stages_roles}
\end{figure*}

Figure \ref{fig:stages_roles} shows the relative participation of each role as a collaborator in the stages of a data science workflow. As motivated in the Related Work section, in this paper, we adopted a six-step view of a reference-model data science workflow, beginning with creating a measurement plan \cite{pine2015politics}, and moving through technical stages to an eventual delivering stage of an analysis or model or working system. Some organizations also perform a check for bias and/or discrimination during the technical development \cite{bellamy2019ai}. However, because that step is not yet accepted by all data science projects and may happen at different stages, we have listed that step separately at the end of the horizontal axis of the stacked bar chart in Figure~\ref{fig:stages_roles}.

\subsubsection{Where do Non-Technical Roles Work?}

The data for Figure \ref{fig:stages_roles} show highly significant differences from one column to the next column ($\chi^2_{48} = 148.777$, p< .001).  

Through a close examination of Figure~\ref{fig:stages_roles}, we found that the degree of involvement by Managers/Executives and by Communicators is roughly synchronized---despite their seeming lack of collaboration patterns as seen in Figures~\ref{fig:roles} and \ref{fig:collaborator_network}. Each of these roles is relatively strongly engaged in the first stage (measurement plan) and the last two stages (evaluate, communicate), but largely absent from the technical work stages (access data, features, model). Perhaps each of these roles is engaged with relatively humanistic aspects of the work, but with \textit{different} and perhaps \textit{unconnected} humanistic aspects for each of their distinct roles.

\subsubsection{Where do Domain Experts Work?}
The involvement of Domain Experts is similar to that of Managers and Communicators, but to a lesser extent. Domain experts are active at every stage, in contrast to Communicators, who appear to drop out during the modeling stage. Domain experts are also more engaged (by self-report) during stages in which Managers have very little engagement. Thus, it appears that Domain Experts are either directly involved in the core technical work, or are strongly engaged in consultation during data-centric activities such as data-access and feature-extraction. They take on even more prominent roles during later stages of evaluating and communicating.

\subsubsection{Where do Technical Roles Work?}
There is an opposite pattern of engagement for the core technical work, done by Engineers/Analysts/Programmers, who are most active while the Managers and Communicators are less involved. 

The degree of involvement by Researchers/Scientists seems to be relatively stable and strongly engaged across all stages. This finding may clarify the ``hub'' results of Figure~\ref{fig:collaborator_network}, which suggested relatively egalitarian collaboration relations. Figure~\ref{fig:stages_roles} suggests that perhaps Researchers/Scientists actively guide the project through all of its stages.

\subsubsection{Who Checks AI Fairness and Bias?}
\label{sec:bias}
The stage of assessment and mitigation of bias appears to be treated largely as a technical matter. Communicators and Managers have minimal involvement. Unsurprisingly, Domain Experts play a role in this stage, presumably because they know more about how bias may creep into work in their own domains. 

\subsubsection{Summary}
From the analyses in this section, we begin to see data science work as a convergence of several analytic dimensions: people in roles, roles in collaboration, and roles in a sequence of project activities. The next section adds a fourth dimension, namely the tools used by data science workers.

\begin{table}[]
 \caption{Categories of data science tools and the number of times each tool was mentioned by respondents.}
\centering
\small
 \begin{tabular}{ l | p{9cm}}
 \toprule
\textbf{Tool Category}  &  \textbf{Tools Mentioned by Respondents} (number of times mentioned) \\ \midrule
asynchronous discussion &  Slack (86), 
email (55), Microsoft Teams (1)
\\ \hline
synchronous discussion &  meeting (13), e-meeting (12), phone (1)
\\ \hline
project management & Jira (8), ZenHub (2), Trello (1)
\\ \hline
code management & GitHub (56), Git (5)
\\ \hline
code & Python (42), R (9), Java (3), scripts (3)
\\ \hline

code editor & 
Visual Studio Code (11), PyCharm (11), RStudio (8), Eclipse (1), Atom (1)
\\ \hline
interactive code environment & 
Jupyter Notebook (66), SQL (6), terminal (4), Google Colab (4)
\\ \hline
software package & Scikit-learn (3), Shiny App (2), Pandas (2)
\\ \hline
analytics/visualization &  SPSS (27), Watson Analytics (22), Cognos (7), ElasticSearch (4), Apache Spark (3), Graphana (2), Tableau (2), Logstash (2), Kibana (1)
\\ \hline
spreadsheet & Microsoft Excel (22), spreadsheets (3), Google Sheets (1)
\\ \hline

document editing & wiki (2), LaTeX (2), Microsoft Word (2), Dropbox Paper (2), Google Docs (1)
\\ \hline
filesharing & Box (43), cloud (5), NFS (2), Dropbox (1), Filezilla (1)
\\ \hline

presentation software & Microsoft Powerpoint (18), Prezi (1)
\\
 \bottomrule
 \end{tabular}
Note: \textbf{code} allows programmers to \textit{write algorithms} for data science. \textbf{code editor} and \textbf{interactive code environment} provide a user experience for writing that code. \textbf{code management} is where the code may be stored and shared. By contrast, \textbf{analytics/visualization} provides ``macro-level'' tools that can invoke entire steps or modular actions in a data science pipeline. 
\label{table:cat}
\end{table}

\subsection{Tooling for Collaboration}
\label{sec:tools}

We asked respondents to describe the tools that they used in the stages of a data science project---i.e., the same stages as in the preceding section. 
We provided free-text fields for them to list their tools, so that we could capture the range of tools used. 
We then collaboratively converted the free-text responses into sets of tools for each response, before iteratively classifying the greater set of tools from all responses into 13 higher-level categories, as shown in Table~\ref{table:cat}.\footnote{We discussed the classification scheme repeatedly until we were in agreement about which tool fit into each category. We postponed all statistical analyses until we had completed our social process of classification.}

\begin{figure*}[]
\includegraphics[width=1\linewidth]{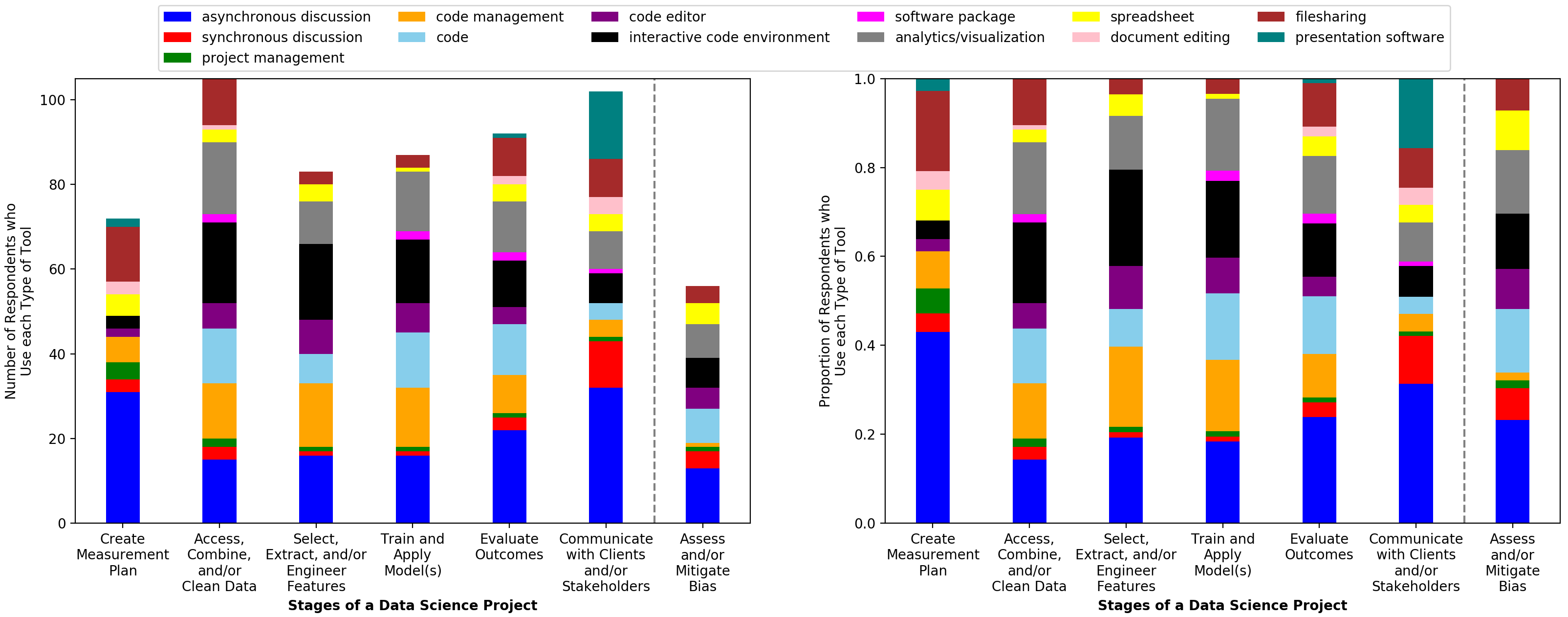}
\caption{The tools used in each stage of a data science project. Note: on the left is the raw count of each tool category for each stage, while on the right, each column is normalized to sum to 1.0.}
\label{fig:stages_tools}
\end{figure*}

When we examined the pattern of tools usage across project stages (Figure \ref{fig:stages_tools}), we found highly significant differences across the project stages ($\chi^2_{72} = 209.519$, p< .001). As above, we summarize trends that suggest interesting properties of data science collaboration:

\subsubsection{Coding and Discussing}
The use of coding resources was as anticipated. Coding resources were used during intense work with data, and during intense work with models. Code may serve as a form of asynchronous discussion (e.g., \cite{brothers1990icicle}): Respondents tended to decrease their use of asynchronous discussion during project stages in which they made relatively heavier use of coding resources.

\subsubsection{Documentation of Work}
We were interested to see whether and how respondents documented their work. Respondents reported some document-editing during the activities leading to a measurement plan. There was also a small use of presentation software, which can of course serve as a form of documentation.\footnote{We consider the use of spreadsheets to be ambiguous in terms of documentation. Spreadsheets function both as records and as discardable scratch-pads and sandboxes. Baker et al. summarize the arguments for treating spreadsheets \textit{not} as documentation, but rather as tools that are \textit{in need of external documentation} (e.g., \cite{davis1996tools}), which is often lacking \cite{baker2006survey}.} The use of these tools returned during the stage of delivery to clients.

\subsubsection{Gaps in Documentation for Feature Engineering} \label{gaps_feature_engineering}
In contrast, we were surprised that there was little use of documents during the phase of feature-extraction and feature-engineering. This stage is an important site for the \textit{design of data}~\cite{feinberg2017design}. The meaning and nature of the data may be changed~\cite{feinberg2017design, muller2019datascience} during this time-consuming step \cite{zoller2019survey,muller2019datascience,guo2011proactive,kandel2011wrangler, rattenbury2017principles}. During this phase, the use of synchronous discussion tools dropped to nearly zero, and the use of asynchronous discussion tools was relatively low. There was relatively little use of filesharing tools. It appears that these teams were not explicitly recording their decisions. Thus, important human decisions may be inscribed into the data and the data science pipeline, while simultaneously becoming invisible~\cite{muller2019datascience, pine2015politics}. The implications for subsequent re-analysis and revision may be severe.

\subsubsection{Gaps in Documentation for Bias Mitigation}
We were similarly surprised that the stage of bias detection and mitigation also seemed to lack documentation, except perhaps through filesharing. We anticipate that organizations will begin to require documentation of bias mitigation as bias issues become more important.

\subsubsection{Summary}
In Section~\ref{do_collaborate}, we showed that data science workers engage in extensive collaboration. Then in Section \ref{stages_roles} we showed that collaboration is pervasive across across all stages of data science work, and that members of data science teams are intensely involved in those collaborative activities. 
By contrast, this section shows gaps in the usage of collaborative tools. We propose that a new generation of data science tools should be created with collaboration ``baked in'' to the design.

\subsection{Collaborative Practices around Code and Data}

Finally, we sought to understand how tool usage by a respondent relates to their practices around code reading, re-use, and documentation, as well as data sharing, re-use, and documentation.
Particularly, if technical team members must collaborate with non-technical team members, then tools and practices to support documentation will be key.

To begin, we clustered the survey respondents into different clusters according to their self-reported tool usage. 
To create a ``tools profile'' for each respondent, we used the questions regarding tool usage, described in Section~\ref{sec:tools}, and summed up all the mentions of each tool from all the open-ended questions on tool usage.
Thus, if a respondent answered only ``GitHub'' for all 7 stages of their data science project, then they would have a count of 7 under the tool ``GitHub'' and a count of 0 elsewhere.

Using the \texttt{k-means} clustering algorithm in the Scikit-learn Python library, we found that \texttt{k}=3 clusters resulted in the highest average silhouette coefficient of 0.254. This resulted in the clusters described in Table~\ref{table:clusters}. 
We only included respondents who had mentioned at least one tool across all the tool usage questions; as the questions were optional, and we experienced some dropout partway through the survey, we had 76 respondents to cluster.

We saw that the respondents in the first cluster (Cluster 0) mentioned using both GitHub and Slack at multiple points in their data science workflow, as well as email and Box to a lesser extent. Given these tools' features for project management, including code management, issue tracking, and team coordination, we characterize this cluster of respondents as \textit{project managed}. 
In contrast, respondents in Cluster 1 mentioned using Jupyter Notebook repeatedly, and only occasionally mentioned other tools; thus we designate the cluster's respondents as using \textit{interactive} tools due to Jupyter Notebook's interactive coding environment.
Finally, Cluster 2 had the most respondents and a longer tail of mentioned tools. However, the tools most mentioned were Python and SPSS; thus, we characterize this cluster of respondents as using \textit{scripted} tools. 
We also noticed that Cluster 2 was predominately made up of self-reported Engineers/Analysts/Programmers at 80\%. Meanwhile, Researchers/Scientists had the greatest prevalence in Cluster 0 and Cluster 1, with 84.2\% and 84.6\%, respectively.

\begin{table}
 \caption{Survey respondents clustered by their self-reported tool usage.}
\centering
\small
 \begin{tabular}{ l | l | l}
 \toprule
\textbf{Respondent Clusters}  & \makecell[l]{\textbf{Number of People}\\\textbf{Per Cluster}} & \makecell[l]{\textbf{Tools Frequently Mentioned}\\(number of times mentioned across questions)} \\ \midrule
0 (\textit{project managed}) & 19 & 
\textbf{GitHub} (86),
\textbf{Slack} (79),
email (47),
Box (26)
\\ \hline
1 (\textit{interactive}) & 13 & \textbf{Jupyter Notebook} (82), GitHub (44),
Slack (22)
\\ \hline
2 (\textit{scripted})& 44 &
\makecell[l]{
\textbf{Python} (50),
SPSS (44),
GitHub (27),
Jupyter notebook\\(27),
Slack (24)}
\\ \bottomrule
 \end{tabular}
\label{table:clusters}
\end{table}

\subsubsection{Reading and Re-using Others' Code and Data}
\begin{figure*}[h]
\includegraphics[width=1\linewidth]{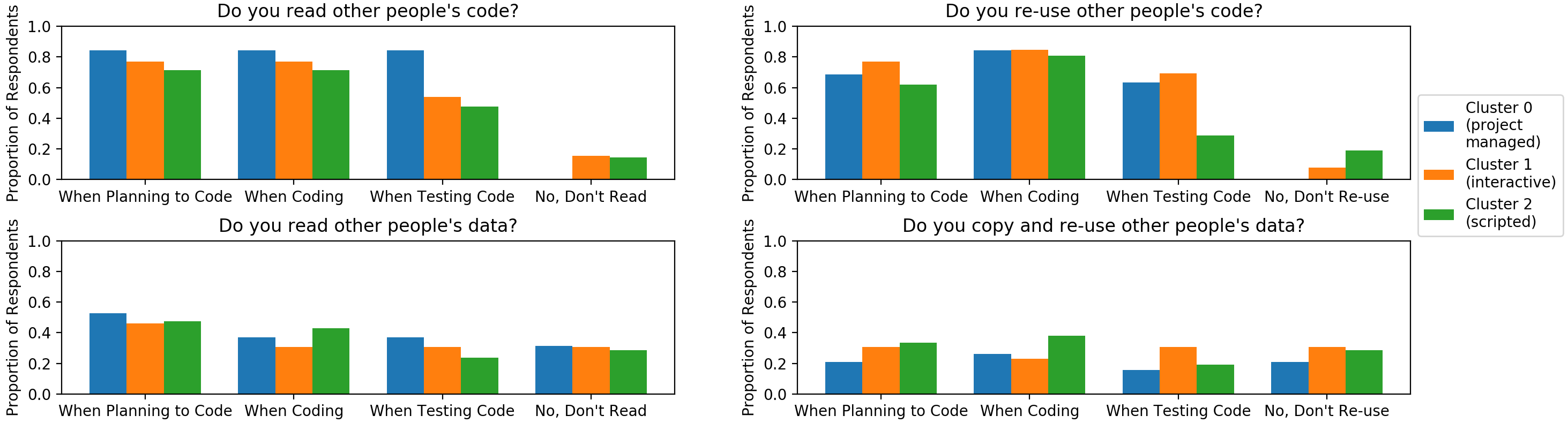}
\caption{Reading and re-use of others' code and data across respondent clusters.}
\label{fig:clusters}
\end{figure*}

In Figure~\ref{fig:clusters}, we report the answers in the affirmative to questions asking respondents whether they read other people's code and data and re-used other people's code and data, separated and normalized by cluster. One finding that stood out is the overall lower levels of collaborative practices around data as opposed to code. This was observed across all three clusters of tool profiles, despite the ability in some tools, such as GitHub, to store and share data.

When comparing across the stages of planning, coding, and testing of code, there were few noticeable differences between clusters except in the stage of \textit{testing} code. Here, we saw that Clusters 1 (interactive) and 2 (scripted) had relatively fewer respondents reading others' code in the testing phase (and Cluster 2 had few respondents re-using others' code in the testing phase). It may be that in an interactive notebook or scripting environment, there is relatively less testing, in contrast to the practice of writing unit tests in larger software projects, and as a result, a relatively lower need for alignment with others' code when it comes to testing. 
We also saw that Cluster 0 (project-managed) had \textit{no} respondents that did not read other people's code or did not re-use other people's code, which suggests that workers in this cluster are coordinating their code with others,
using tools like GitHub and Slack.

\subsubsection{Expectations Around One's Own Code and Data Being Re-used}


\begin{table}[t]
    \caption{Expectations around one's own code and data being re-used.}
    \small
    \centering
    \begin{tabular}{ l | r | r | r | r}
    \toprule
  &  \makecell{Cluster 0\\(\textit{Project managed})} & \makecell{Cluster 1\\ (\textit{Interactive})} & \makecell{Cluster 2\\ (\textit{Scripted})} & All \\
    \midrule
    Expect that their \textbf{code} will be re-used  & 68.4\% & 84.6\% & 80.9\% & 78.8\% \\
     \midrule
   Expect that their \textbf{data} will be re-used  & 73.7\% & 46.1\%  & 50\% & 59.6\% \\
    \bottomrule
    \end{tabular}
    \label{tab:reuse}
\end{table}

In Table~\ref{tab:reuse}, we report on respondents answers to their \textit{expectations} for how their own code or data will be used by others. Respondents were more likely to state that they expected others to re-use their code as opposed to their data.
In the case of code re-use, peoples' expectations were slightly lower for respondents in Cluster 0 and slightly higher for respondents in the other clusters, though this was not significant.
We also saw that the expectation that data would be re-used was more prevalent in Cluster 0 while relatively low in Cluster 1 and 2.
This may be because the native features for displaying structured data or coordinating the sharing of data, such as using version control, are more rudimentary within tools like Jupyter Notebook, although a few recent works have developed prototypes examined in a lab environment~\cite{kery2019towards,wang2019data,wang2020callisto}.

\begin{table}
    \caption{Code and data documentation practices according to each cluster.}
    \small
    \centering
    \begin{tabular}{ l | l| r | r | r}
    \toprule
  & \textbf{Documentation Practice} &  \makecell{Cluster 0\\(\textit{Project managed})} & \makecell{Cluster 1\\ (\textit{Interactive})} & \makecell{Cluster 2\\ (\textit{Scripted})}\\
    \midrule
    \multirow{4}{*}{ Code } &
     In-line comments & 100\% & 84.6\% & 90.5\%\\
   & Longer blocks of comments in the code & 68.4\% & 30.8\% & 38.1\%\\
  &  Markdown cells in notebooks & 63.2\% & 92.3\% & 38.1\%\\
  &  External documents & 63.2\% & 61.5\% & 28.6\%\\
     \midrule
   \multirow{4}{*}{ Data } & Column labels & 66.7\% & 63.6\% & 75\% \\
  &  Data dictionary & 50\% & 63.6\% & 40\%\\
  &  Extra text in JSON schema (or similar) &
    27.8\% & 18.2\% & 15\%\\
  &  External documents & 77.8\% & 45.5\%  & 50\%\\
    \bottomrule
    \end{tabular}
    \label{tab:practices_cluster}
\end{table}

\subsubsection{Code and Data Documentation Practices}

In Table~\ref{tab:practices_cluster}, we show respondents' answers to how they document their code as well as their data, broken down by cluster. 
Overall, we see higher rates of self-reported code documentation in Cluster 0 (project managed) and 1 (interactive) compared to 2 (scripted). 
For instance, 100\% of members of Cluster 0 said they used in-line comments to document their code. 
Cluster 2 also had high rates of using in-line comments, though other practices were infrequently used. 
Unsurprisingly, the use of markdown cells in notebooks was most prevalent in Cluster 1 (interactive), while longer blocks of comments in the code was least used (30.8\%) likely because markdown cells perform that function. 
A lack of code documentation besides in-line comments in Cluster 2 suggests that it may be more difficult for collaborators to work with code written by members of this cluster. 
We note that this is not due to a low expectation within Cluster 2 that code would be re-used.

We found that data science workers overall performed less documentation when it comes to data as opposed to code, perhaps due to their perceptions around re-use. Even something basic like adding column labels to data sets was not performed by a third to a quarter of members of each cluster, as shown in Table~\ref{tab:practices_cluster}. Instead, the most prevalent practice within any of the clusters was the use of external documents by Cluster 0 (project managed) at 77.8\%. While external documents allow data set curators to add extensive documentation about their data, one major downside is that they are uncoupled---there is little ability to directly reference, link to, or display annotations on top of the data itself. This may lead to issues where the documentation can be lost, not noticed by a collaborator, or misunderstood out of context.

\subsubsection{Summary}
In this section, we examined the collaborative practices of data science workers in relation to the kinds of tools they use. Through clustering, we identified three main ``tools profiles''. The first makes heavy use of GitHub and Slack and is relatively active in reading other people's code, re-using other people's code, expecting that others would use one's code and data, and documenting code.
Out of all the three clusters, workers using this tool profile seem to have the healthiest collaborative practices. However, even this cluster has relatively low rates of collaboration and documentation around data.

The second cluster primarily uses Jupyter Notebook for data science work. While people in this cluster were generally active in code collaboration and code documentation, we notice a lower rate of reading others' code while testing one's code as well as a low expectation that one's data would be re-used. 

The third cluster had a greater variety of tool usage but more emphasis on writing scripts in Python or SPSS.
This cluster had low rates of code documentation outside of in-line comments, signaling potential difficulties for non-technical collaborators.

\section{Discussion}

We began our Results by asking ``Do data science workers collaborate?'' The answer from this survey dataset is clearly ``yes.'' These results are in agreement with prior work~\cite{grappiolo2019semantic, hou2017hacking, passi2017data, stein2017make, borgman2012s, mao2019, viaene2013data, wang2019humanai}. In this paper, we provide a greater depth of collaboration information by exploring the interactions of team roles, tools, project stages, and documentation practices. One of our strong findings is that people in most roles report extensive collaboration during each stage of a data science project. These findings 
suggest new needs among data science teams and communities, and encourage us to think about a new generation of ``collaboration-friendly'' data science tools and environments.

\subsection{Possible Collaborative Features}


\subsubsection{Provenance of data}

We stated a concern earlier that there seemed to be insufficient use of documentation during multiple stages of data science projects and fewer practices of documentation for data as opposed to code. 
Partly this may be due to a lack of expectations that one's data will ever be re-used by another.
In addition, there are now mature tools for collaborating on code due to over a decade of research and practice on this topic in the field of software engineering~\cite{treude2009empirical,dabbish2012social}; however, fewer tools exist for data and are not yet widely adopted. 
The absence of documentation may obscure the source of datasets as well as computations performed over datasets in the steps involving data cleaning or transformation. The problem can be compounded if there is a need to combine datasets for richer records. 
When teams of data science workers share data, then the knowledge of one person may be obscured, and the organizational knowledge of one team may not be passed along to a second team. Thus, there is a need for a method to record data provenance. 
A method for embedding this information within the data themselves would be a superior outcome as opposed to within external documents.
As one example, the DataHub project~\cite{bhardwaj2014datahub} replicates GitHub-like features of version control and provenance management but for datasets. In a similar vein, the ModelDB project provides version control and captures metadata about machine learning models over the course of their development~\cite{vartak2016m}.
Beyond provenance captured automatically, there needs to be ways for collaborators to record discussions and decisions made with each transformation.

\subsubsection{Provenance of code}
There also remain subtle issues in the provenance of code. At first, it seems as if the reliance of data science on code packages and code libraries should obviate any need for documentation of code in data science. However, in Section~\ref{gaps_feature_engineering}, we discussed the invisibility of much of the work on feature extraction and feature engineering. The code for these activities is generally not based on a well-known and well-maintained software package or product. If this code becomes lost, the important knowledge about the nature and meaning of the features \cite{muller2019human, muller2019datascience} may also be lost. 

However, lack of motivation to document lower-level decision-making may be a limiting factor towards stronger documentation practices, particularly in an ``exploration'' mindset. In the software engineering realm, tools have been proposed to support more lightweight ways for programmers to externalize their thought processes, such as social tagging of code~\cite{storey2006shared} or clipping rationales from the web~\cite{liu2019unakite}. Other tools embed and automatically capture context while programmers are foraging for information to guide decisions, such as search~\cite{brandt2010example} and browsing history~\cite{hartmann2011hypersource,fourney2013enhancing}. 
Similar ideas could be applied in the case of data science, where users may be weighing the use of different code libraries or statistical methods. Other decisions such as around feature engineering may result from conversations between team members~\cite{Coding_team:Park:2018:PPL:3266037.3266098} that then could be linked in the code.

In addition, we noticed a drop-off in collaborative code practices when it came to testing already-written code. This has important implications for developing standards around testing for data and model issues of  bias, which will only be more important in years to come. 
Thus, preserving the provenance of code may also be important to keep the data processing steps transparent and accountable. 

\subsubsection{Transparency}
More broadly, data science projects may inadvertently involve many assumptions, improvisations, and hidden decisions.
Some of these undocumented commitments may arise through the assumption that everyone on the team shares certain knowledge---but what about the next team that ``inherits'' code or data from a prior project? As we just noted, this kind of transparent transmission of knowledge may be important with regard to the design of features~\cite{feinberg2017design, muller2019datascience}. It can also be important for the earlier step of establishing a data management plan, which may define, in part, what qualifies as data in this project~\cite{pine2015politics}.

We advocate to make invisible activities more visible---and thus discuss-able and (when necessary) debatable and accountable. This argument for transparency is related but distinct from the ongoing ``Explainable AI'' (XAI) initiative---XAI emphasizes using various techniques (e.g., visualization~\cite{karl2020}) and designs to make machine learning algorithms understandable by non-technical users~\cite{amershi2019guidelines,JaimieDrozdal2020,heer2019agency, liao2020questioning,zhang2020effect}, whereas we argue for the explanation of decisions among the various data science creators of machine learning algorithms.
Recent work in this space similarly argues for more documentation to improve transparency, as well as greater standardization around documentation~\cite{mitchell2019model,gebru2018datasheet},  particularly important when it comes to publicly-released datasets and models.

\subsection{Collaborating with Whom? and When?}
These concerns for provenance and transparency may be important to multiple stakeholders. 
Team members are obvious beneficiaries of good record-keeping. In the language of value sensitive design~\cite{friedman2013value}, team members are \textit{direct stakeholders}---i.e., people who directly interact with data science tools in general, and the project's code in particular. Again using concepts from value sensitive design, there are likely to be multiple \textit{indirect stakeholders}---i.e., people who are affected by the data science system, or by its code, or by its data. 

\subsubsection{Indirect Stakeholders}
For a data science project, indirect stakeholders might be future project teams. These peers (or future peers) would benefit from understanding what decisions were made, and how data were defined~\cite{pine2015politics} and transformed~\cite{feinberg2017design, muller2019datascience}. 
For data science projects that affect bank loans~\cite{bruckner2018promise, o2016weapons} prison sentences~\cite{picardbeyond}, or community policing~\cite{verma2018confronting}, the public are also indirect stakeholders as they worry about the possibility of inequitable treatment or faulty data. 
Finally, another beneficiary of provenance and transparency is one's own future self, who may return to a data science project after a year or two of other engagements, only to discover that the team has been dispersed, personal memories have faded, and the project needs to be learned like any unfamiliar data science resource.

\subsubsection{``Imbalanced'' Collaboration}
\label{sec:discussion_collaboration}
In Section \ref{sec:collaboration}, we observed that there is a mismatch around perceived collaborations between different roles. For example, Communicators believed they collaborate a lot with Managers/Executives, but the Managers/Executives perceived they collaborated the least with Communicators. 
This result is the normalized proportions of the reported collaborations from each role in Figure~\ref{fig:roles}, so it is possible that Managers/Executives may collaborate a lot with all other roles and the collaboration with Communicators has the smallest proportion among these collaborations.

We speculate that the collaborations reported by our informants may have been highly directional. Communicators may have received information from other roles---or may simply have read shared documents or observed meetings---to find the information that they needed to communicate. Their role may have been largely to \textit{receive} information, and they are likely to have been aware of their dependencies on other members of the team. By contrast, the other roles may have perceived Communicators as relatively passive team members. These other roles may have considered that they themselves received little \textit{from} the Communicators, and may have down-reported their collaborations accordingly.

Different roles also reported different intra-role collaboration patterns in Section~\ref{sec:intra_collaboration}. These patterns suggest that the people in these roles may have different relationships with their own communities of practice \cite{duguid2005art, wenger2011communities}. There may be stronger peer communities for each of Engineers, Researchers, and Communicators, and there may be weaker peer communities for each of Managers and Domain Experts. It may be that Domain Experts are focused within their own domain, and may not collaborate much with Domain Experts who work in other domains. It may be that Managers/Executives focus on one data science project at-a-time, and do not consult with their peers about the technical details within each project.


\subsection{AI Fairness and Bias}

The detection, assessment, and mitigation of bias in data science systems is inherently complex and multidisciplinary, involving expertise in prediction and modeling, statistical assessments in particular domains, domain knowledge of an area of possible harms, and aspects of regulations and law. There may also be roles for advocates for particular affected groups, and possibly advocates for commercial parties who favor maintaining the status quo. 

In these settings, a data science pipeline becomes an object of contention. Making sense of the data science pipeline requires multiple interpreters from diverse perspectives, including adversarial interpreters \cite{friedman2013value}. All of the issues raised above regarding provenance and transparency are relevant.

Our result confirmed that in data science collaborations, there are activities around AI fairness and bias detection and mitigation happening along the data science workflow in Section \ref{sec:bias}, and it appears to be treated largely as a technical matter. For example, data scientists and engineers are involved in the process as they follow up with the latest technical algorithms on how to detect bias and fix it. Our results also suggest that Domain Experts also plays a role in the Bias detection and Mitigation process, presumably because they know more about how bias may creep into work in their own domains. 

However, we did not see much involvement from Communicators and Managers/Executives. This is surprising, as Communicators and Managers are the ones who may know the most about policy requirements and worry the most about the negative consequences of a biased AI algorithm. We speculate that Managers may become more involved in this stage in the future, as bias issues become more salient in industry and academia \cite{abiteboul2016data, garcia2016racist, hajian2016algorithmic}.

\subsection{Limitations and Future Directions}
Our survey respondents were all recruited from IBM---a large, multinational technology company---and their views may not be fully representative of the larger population of professionals working in the data science related projects. 

One example of how our results might be skewed comes from the fact that almost all of our respondents worked in small teams, typically with 5 or 6 collaborators in a team. While this number is consistent to what previous literature reported (e.g., Wang et al.~reported 2-3 data scientists in a team~\cite{wang2019humanai}, and our work also counts managers, communicators, researchers, and engineers), in other contexts the size of data science teams may vary. Also, due to the fact that all these respondents are from the same company, their preference in selecting tools and how to use these tools may be dominated by the company culture. The findings may be different if we study data science teams' collaborative practices in different scenarios, such as in offline data hackathons~\cite{hou2017hacking}.

Another limitation is that our findings are based on self-reported data using an online survey. Despite this research method's power of covering a broader user population, it is also known that survey respondents may have bias in answering those behavioral questions. We see this rather as a new promising research direction than limitation, and we look forward to conducting further studies with Contextual Inquiry~\cite{siek2014field}, Participatory Analysis \cite{muller2001layered}, or Value Sensitive Design~\cite{friedman2013value} to observe and track how people actually behave in a data science team collaboration.

We should also note that data science teams may not always appreciate the proposed features that increase transparency and accountability of each team member's contribution, as they may have negative effects. In the co-editing activities enabled by Google Doc-like features, writers sometimes do not want to have such high transparency~\cite{Wang:2015:DVC:2702123.2702517}. Thus, we need additional user evaluations of collaboration features before deploying them into the real world.

\section{Conclusion}

In this paper, we presented results of a large-scale survey of data science workers at a major corporation that examined how data science workers collaborate. We find that not only do data science workers collaborate extensively, they perform a variety of roles, and work with a variety of stakeholders during different stages of the data science project workflow.  We also investigated the tools that data scientists use when collaborating, and how tool usage relates to collaborative practices such as code and data documentation. From this analysis, we present directions for future research and development of data science collaboration tools.

In summary, we hope we have made the following contributions:
\begin{itemize}
    \item The first large in-depth survey about data science collaborative practices, and the first large study to provide roles-based analyses of collaborations.
    \item The first large-scale study of data science activities during specific stages of data science projects.
    \item The first analysis of collaborative tools usage across the stages of data science projects.
    \item The first large-scale analysis of documentation practices in data science.
\end{itemize}

\received{October 2019}
\received[revised]{January 2020}
\received[accepted]{January 2020}

\begin{acks}
We appreciate all the survey respondents for participating in our online survey. This work is generously supported by MIT-IBM Watson AI Lab under the ``Human-in-the-loop Automated Machine Learning'' project. 
\end{acks}

\bibliographystyle{ACM-Reference-Format}
\bibliography{sample-journal}

\end{document}